\documentclass[
reprint,
superscriptaddress,
amsmath,amssymb,
aps,
pra,
longbibliography,
]{revtex4-1}
\usepackage{float}
\usepackage{xcolor}

\usepackage{graphicx}
\usepackage{dcolumn}
\usepackage{bm}
\usepackage[colorlinks=true, pdfstartview=FitV, linkcolor=blue, citecolor=blue, urlcolor=blue]{hyperref}
\usepackage{dsfont}

\usepackage[separate-uncertainty=true]{siunitx}
\usepackage[siunitx]{circuitikz}

\graphicspath{{figures/}}

\begin{document}

\preprint{APS/123-QED}

\title{Deterministic and stochastic sampling of two coupled Kerr parametric oscillators}

\author{Gabriel Margiani}
\affiliation{Laboratory for Solid State Physics, ETH Z\"{u}rich, CH-8093 Z\"urich, Switzerland.}
\author{Javier del Pino}
\affiliation{Institute for Theoretical Physics, ETH Z\"{u}rich, CH-8093 Z\"urich, Switzerland.}
\author{Toni L. Heugel}
\affiliation{Institute for Theoretical Physics, ETH Z\"{u}rich, CH-8093 Z\"urich, Switzerland.}
\author{Nicholas E. Bousse}
\affiliation{Departments of Mechanical Engineering, Stanford University, Stanford, California 94305, USA}
\author{Sebastián Guerrero}
\affiliation{Laboratory for Solid State Physics, ETH Z\"{u}rich, CH-8093 Z\"urich, Switzerland.}
\author{Thomas W. Kenny}
\affiliation{Departments of Mechanical and Electrical Engineering, Stanford University, Stanford, California 94305, USA}
\author{Oded Zilberberg}
\affiliation{Department of Physics, University of Konstanz, D-78457 Konstanz, Germany.}
\author{Deividas~Sabonis}
\affiliation{Laboratory for Solid State Physics, ETH Z\"{u}rich, CH-8093 Z\"urich, Switzerland.}
\author{Alexander Eichler}
\affiliation{Laboratory for Solid State Physics, ETH Z\"{u}rich, CH-8093 Z\"urich, Switzerland.}

\date{\today}

\begin{abstract}
The vision of building computational hardware for problem optimization has spurred large efforts in the physics community. In particular, networks of Kerr parametric oscillators (KPOs) are envisioned as simulators for finding the ground states of Ising Hamiltonians. It was shown, however, that KPO networks can feature large numbers of unexpected solutions that are difficult to sample with the existing deterministic (i.e., adiabatic) protocols. In this work, we experimentally realize a system of two classical coupled KPOs, and we find good agreement with the predicted mapping to Ising states. We then introduce a protocol based on stochastic sampling of the system, and we show how the resulting probability distribution can be used to identify the ground state of the corresponding Ising Hamiltonian. This method is akin to a Monte Carlo sampling of multiple out-of-equilibrium stationary states and is less prone to become trapped in local minima than deterministic protocols.
\end{abstract}

	\maketitle

The Kerr parametric oscillator (KPO) is a nonlinear system whose potential energy is modulated at a frequency $f_p$ close to twice its resonance frequency, $f_p \approx 2f_0$~\cite{Ryvkine_2006, Mahboob_2008, Wilson_2010, Eichler_2011_NL, Leuch_2016, Gieseler_2012, Lin_2014, Puri_2017, Eichler_2018, Nosan_2019, Frimmer_2019, Grimm_2019, wang_2019, Puri_2019_PRX, Miller_2019_phase, yamaji_2022}. When the modulation depth $\lambda$ exceeds a threshold $\lambda_\mathrm{th}$, the system features two stationary oscillation solutions. The solutions have a frequency $f_p/2$, an amplitude $X$ determined by $\lambda$ relative to the Kerr nonlinearity, and phases that differ by $\pi$. These `phase states' can be mapped to the two states $\sigma \in \{-1,1\}$ of an Ising spin. Building on that analogy, it was proposed that networks of KPOs can be utilized to simulate the ground state of coupled spin ensembles, as captured by the Ising model Hamiltonian~\cite{Ising_1925}:
\begin{align*}
    H_\mathrm{Ising} = -\sum_{i,j} K_{i,j}\sigma_i\sigma_j\,,
\end{align*}
where $K_{i,j}$ is the coupling coefficient between two spins with states $\sigma_{i,j}$. Interestingly, finding this ground state is equivalent to many computational problems that are nearly intractable with conventional computers~\cite{mohseni2022ising}, such as the number partitioning problem~\cite{Nigg_2017}, the MAX-CUT problem~\cite{Inagaki_2016_Science, Goto_2019}, and the famous traveling salesman problem~\cite{Lucas_2014}.

Various physical implementations have been proposed or realized as ``Ising solvers''~\cite{Gottesman_2001, Devoret_2013, Mahboob_2016, Inagaki_2016, Goto_2016, Puri_2019_PRX,Bello_2019,Okawachi_2020}. A well known example is the Coherent Ising Machine, a network of degenerate optical parametric oscillators (DOPOs) that are coupled through a programmable electronic feedback element~\cite{Wang_2013,Inagaki_2016_Science,Mahboob_2016,Yamamoto_2017,Yamamura_2017,Bello_2019} (note that DOPOs differ from KPOs in that their amplitude is not determined by their Kerr  nonlinearity but rather by two-photon loss). The feedback breaks the energy conservation of the network and imparts dissipative coupling (mutual damping) between the oscillators. As a consequence, different network configurations corresponding to different Ising solutions become stable at different driving thresholds. The optimal solution is assumed to possess the lowest threshold and therefore to appear as the solution when driving the network.

A different line of investigation focuses on energy-conserving, bilinear coupling between KPOs~\cite{Goto_2016,Puri_2017_NC,Nigg_2017,Goto_2018,Dykman_2018,Rota_2019,Heugel_2022}. In this works, the coupled oscillator solutions can be approximated as decoupled normal modes with split resonance frequencies. In contrast to the case of dissipative coupling, the lowest threshold that is encountered when ramping up the driving strength depends here on the detuning $\Delta$ between the external drive and the resonance frequency. This control parameter opens up the possibility for specific protocols to find different solutions~\cite{Goto_2016,Puri_2017_NC,Nigg_2017}. It was experimentally demonstrated, however, that the combination of nonlinearities and strong bilinear coupling can give rise to a rich solution space beyond that of a simple Ising model~\cite{Heugel_2022}. Careful validation and testing of small systems is therefore important before larger networks can be understood and operated correctly.

In this paper, we experimentally test the validity of the Ising analogy for a system of two classical coupled KPOs.
In a first step, we apply an adiabatic ramping protocol to find one particular solution for each selected combination of $\Delta$ and $\lambda$ in a deterministic way. In a second step, we use strong force noise to explore the solution space of the system: this method is based on transitions between different stationary KPO solutions~\cite{Dykman_1998,Chan_2007,Chan_2008,Mahboob_2014_2,margiani2021fluctuating}, and it allows for the visualization of a probability distribution for all accessible states. Such ``stochastic sampling'' presents a way of quantifying the occupation probability of each solution, and thus selecting the optimal state. Surprisingly, we find that the oscillation state corresponding to the expected Ising ground state has not the highest but the \textit{lowest} occupation probability over the entire parameter space. We reconcile this result with theory and predict how the method can be used for larger networks.

\begin{figure}[t]
    \includegraphics[width=\columnwidth]{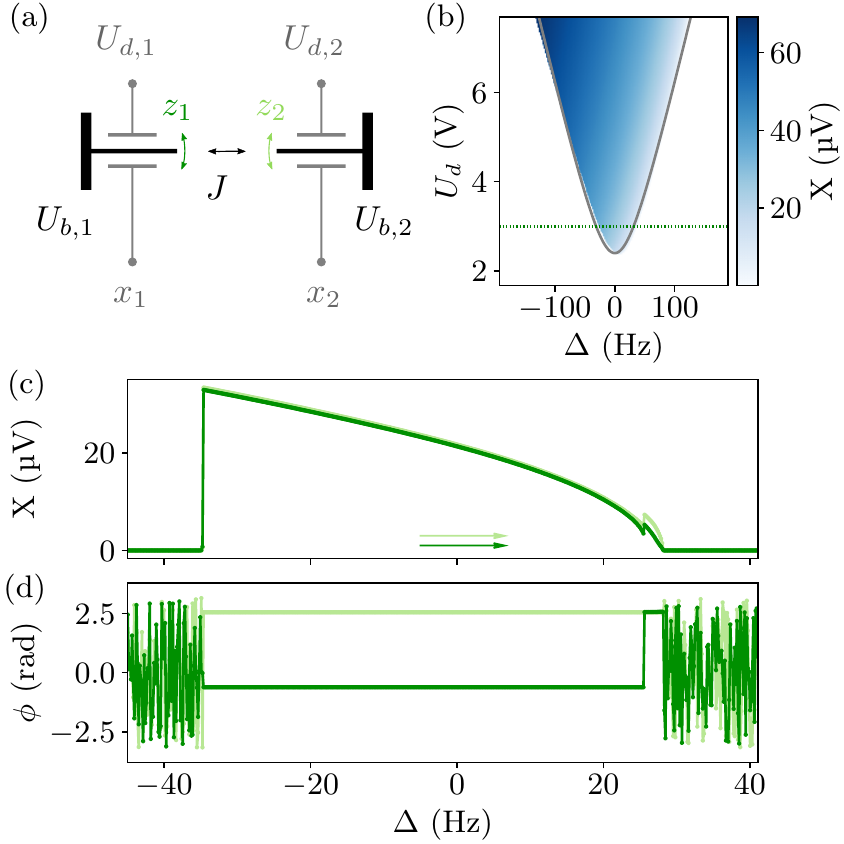}
    \caption{\textbf{Device and basic characterization.} (a)~Simplified schematic of our setup: Two mechanical resonators are coupled with  a strength $J$ via their common substrate. The devices are charged (and tuned) with bias voltages $U_{b,i}$ and driven by $U_{d,i}$. Their displacement $z_i$ capacitively translates into  a voltage $x_i$ that is read out. (b)~The Arnold tongue of a single KPO indicates the region in a space spanned by $\Delta$ and $U_\mathrm{d}\propto\lambda$ where the zero-amplitude state becomes unstable and the resonators respond with a finite amplitude $X$. The solid gray line indicates the theoretical threshold $U_\mathrm{th}$ (see Appendix~\ref{sec:apDevices}). (c)~Amplitude and (d)~phase of the two resonators in a frequency sweep from low to high frequencies (direction of arrows) with $U_\mathrm{d} = \SI{3}{\volt}$; cf. the green dashed line in (b). Close to $\Delta = \SI{25}{\hertz}$, one of the KPOs (dark green) flips its phase while the other (bright green) remains in the same phase state. The system's state changes from antisymmetric to symmetric at this point. Outside of the Arnold tongue, the phases are undefined. Note that for the opposite sweep direction, the state change does not occur. Instead, the system follows the symmetric state as long as it remains stable.}
    \label{fig:fig1}
\end{figure}

\begin{figure*}[t]
    \includegraphics[width=\textwidth]{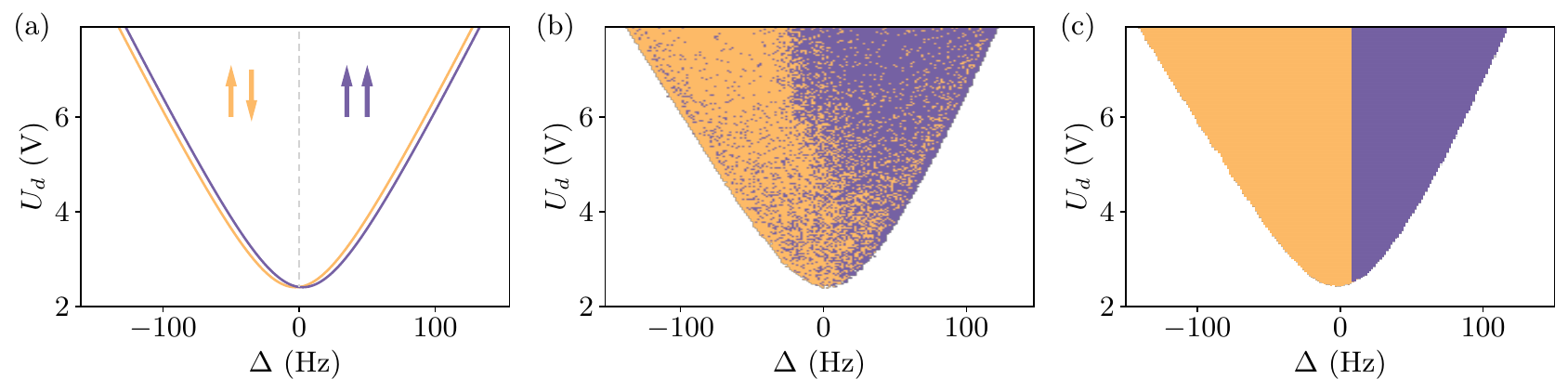}
    \caption{\textbf{Deterministic sampling of the coupled system.} (a)~Schematic shapes (with exaggerated splitting) of the two overlapping Arnold tongues centered at $f_{A,S} = f_0 \pm\Delta_f/2$. Each tongue has the boundaries $f_\pm^2 =  f_{A,S}^2\left(1\pm \sqrt{\frac{\lambda^2}{4}-\frac{1}{Q^2}}\right)$. The tongue for the antisymmetric and symmetric normal mode are shown in orange and purple, respectively. $\Delta = 0$ is shown as a dashed grey line and arrows indicate the solution with the lowest threshold for positive and negative $\Delta$. (b)~Measured response to a parametric drive initialized with fixed strength $U_\mathrm{d}$ and detuning $\Delta$. For each drive-frequency pair, the parametric drive was activated and the system was given approx. \SI{400}{\milli\second} to settle into a solution before the result was recorded with an integration time of \SI{16}{\milli\second}. Afterwards, the drive was switched off for \SI{400}{\milli\second} before moving on to the next coordinate. Symmetric (purple) and antisymmetric (orange) solutions were identified by comparing the measured phases of the two KPOs, while the shape of the Arnold tongue was extracted from their amplitudes. (c)~Measured response to a parametric drive initialized with detuning $\Delta$ and a strength that is slowly ramped upwards from $U_\mathrm{d}=0$ with no initial oscillation (white). The color coding is the same as in (b).}
    \label{fig:fig2}
\end{figure*}

Our system comprises two microelectromechanical resonators (MEMS) made from highly-doped single-crystal silicon. Both resonators are fabricated on the same chip in a wafer-scale encapsulation process \cite{yang2016unified}, and they have the shape of double-ended tuning forks with branches \SI{200}{\micro\metre} long and \SI{6}{\micro\metre} thick; see Fig.~\ref{fig:fig1}(a) and Appendix~\ref{sec:apDevices}. They have resonance frequencies of roughly $f_0 \approx \SI{1.124}{\mega\hertz}$ and quality factors of $Q \approx 13500$.  Bias voltages $U_b$ can be used to fine-tune the resonator frequencies by a few \si{\kilo\hertz} and induce negative Kerr-nonlinear coefficients of $\beta \approx \SI{-63.2e17}{\per\volt\squared\per\second\squared}$ due to the nonlinear electrostatic forces between the biased tuning fork and the electrodes next to it~\cite{agarwal2008study}. Those electrodes capacitively transduce the motion into electrical signals that are measured with a Zurich Instruments HF2LI lock-in amplifier. The capacitive driving and measurement allows us to write effective equations of motion~\cite{Nosan_2019} as
\begin{align*}
	&\ddot{x}_i + \omega_0^2\left[1-\lambda\cos\left(2\pi f_p t\right)\right]x_i + \beta x_i^3 + \gamma \dot{x}_i - Jx_j = U_{\xi,i}\,,
\end{align*}
where $\omega_0/2\pi = f_0$, $\gamma = \frac{\omega_0}{Q}$, $x_{i}$ is the measured voltage signal of resonator $i$, $J$ quantifies the coupling to resonator $j$, and $U_{\xi,i}$ indicate uncorrelated white noise sources. The electrical tuning effect allows us to parametrically modulate (drive) the resonator potentials at frequency $f_p$~\cite{margiani2021fluctuating}. The required oscillating driving voltage is $U_\mathrm{d}=\lambda U_\mathrm{th}^0 Q/2$, where $U_{\mathrm{th}}^0 \approx \SI{2.4}{\volt}$ is the measured parametric threshold voltage on resonance. As a function of detuning $\Delta = f_p/2 - f_0$, the driving threshold for parametric oscillation, $U_\mathrm{th}$, is described by a so-called `Arnold tongue', see Fig.~\ref{fig:fig1}(b). Outside the Arnold tongue, a resonator is stable at zero amplitude, while inside the tongue the zero-amplitude solution becomes unstable and the resonator oscillates at $f_p/2$ with a finite effective amplitude $X$ in one out of two possible phase states.

The resonators are mechanically coupled via their common substrate~\cite{miller2022influence}. We calibrate the coupling strength from the normal-mode splitting and find $\Delta_f = \SI{-2.6(3)}{\hertz}$, corresponding to a coupling coefficient $J = 4\pi^2\Delta_f f_0 = -\SI{113(13)e6}{\hertz\squared}$ between the two resonators; see Appendix~\ref{sec:apDevices} for details. Even though the coupling is weak, $|\Delta_f| \ll f_0/Q$, we can use a normal-mode basis of antisymmetric and symmetric oscillations to describe our system in the following.

When both resonators are operated as KPOs with a parametric drive voltage $U_{\mathrm{d}} > U_\mathrm{th}$, each of them selects one of its two phase states. The resonators can respond either in the same (symmetric) or in opposite (antisymmetric) phases. Which of those two solutions is preferred depends on the signs of $J$ and $\Delta$. In Figs.~\ref{fig:fig1}(c) and \ref{fig:fig1}(d), we show the experimental results of sweeping the driving frequency from negative to positive $\Delta$ at a fixed $U_{\mathrm{d}}$. The system first rings up into the antisymmetric state at $\Delta = \SI{-35}{\hertz}$ before it flips to a symmetric configuration close to $\Delta = \SI{25}{\hertz}$.

The ordering observed in Figs.~\ref{fig:fig1}(c) and \ref{fig:fig1}(d) reflects the fact that for $J < 0$, the antisymmetric normal mode has a lower eigenfrequency than the symmetric mode. We can  therefore expect to find separate normal-mode Arnold tongues for symmetric and antisymmetric oscillations with a splitting in frequency; see Fig.~\ref{fig:fig2}(a)~\cite{Heugel_2019_TC,Heugel_2022}. This splitting has important consequences for the driven system: when the drive is suddenly activated at a specific detuning $\Delta$ above threshold, the KPOs should preferentially select the normal-mode oscillation state with the lowest threshold. We experimentally test this prediction in Fig.~\ref{fig:fig2}(b) by measuring the chosen state once for each pixel individually, and we find very good agreement with the schematic in Fig.~\ref{fig:fig2}(a). Note the narrow regions on the left and right boundaries where only one state can be activated. These regions are a direct confirmation of the normal-mode splitting.

\begin{figure}[t]
    \includegraphics[width=\columnwidth]{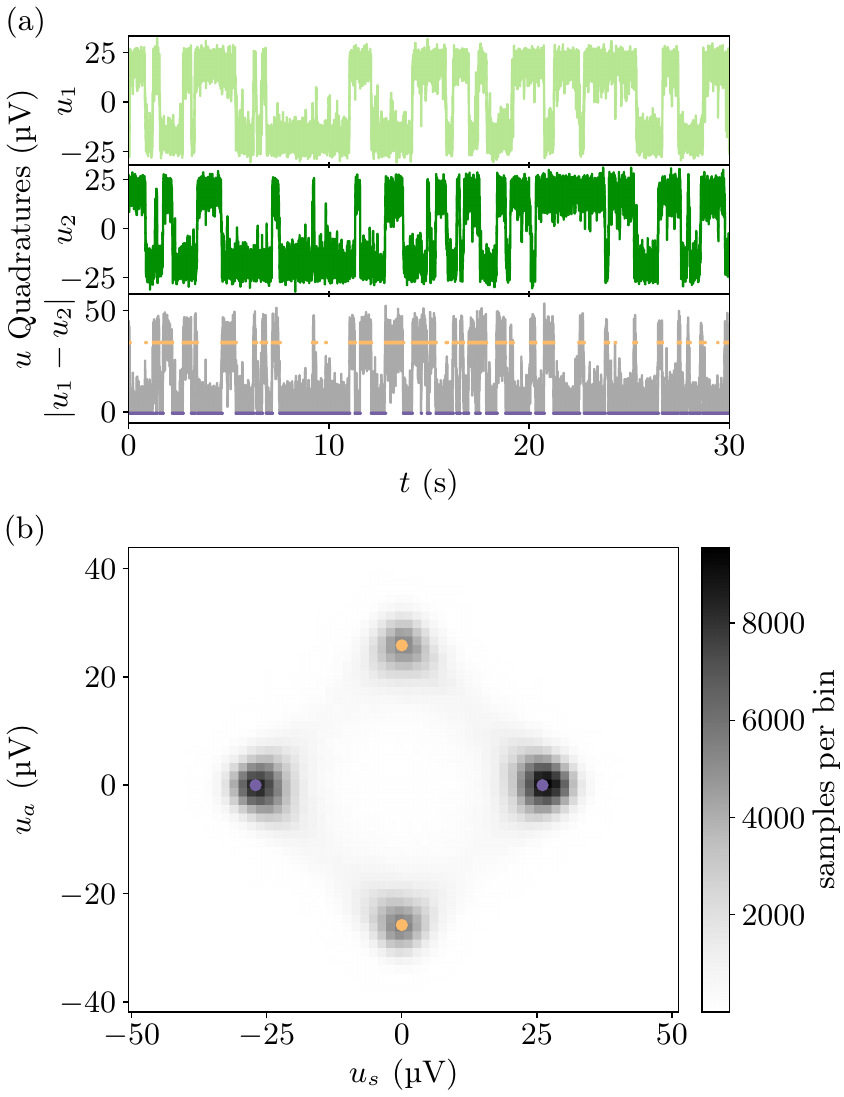}
    \caption{\textbf{Stochastic sampling of the coupled system.} (a)~Extract from a measured sample path of both KPOs under the influence of applied force noise. Switches between the quasistable states are visible as jumps from $-\SI{25}{\milli\volt}$ to $+\SI{25}{\milli\volt}$ and vice versa. Only the $u$ quadratures and their difference are shown for simplicity. A high difference indicates that the system is in an antisymmetric state (orange dots), while differences close to zero correspond to a symmetric state (purple dots). (b) Representation of the entire sample path in a ``super phase space'' spanned by $u_s = (u_1 + u_2)/\sqrt{2}$ and $u_a = (u_1 - u_2)/\sqrt{2}$. Orange and purple dots mark the antisymmetric ($u_s \approx 0$) and symmetric ($u_a \approx 0$) phase state configurations, respectively. The graph also shows how switches mostly involve a change in symmetry from symmetric to antisymmetric or vice versa, while symmetry-preserving transitions through the origin of the plot are rare. We use $U_\mathrm{d} = \SI{3}{\volt}$ and $\Delta = 0$. The total measurement time was \SI{6}{\min}, measured with an integration time of \SI{138}{\micro\second} at 3597 samples per second.} 
    \label{fig:fig3}
\end{figure}

\begin{figure}[h!]
    \includegraphics[width=\columnwidth]{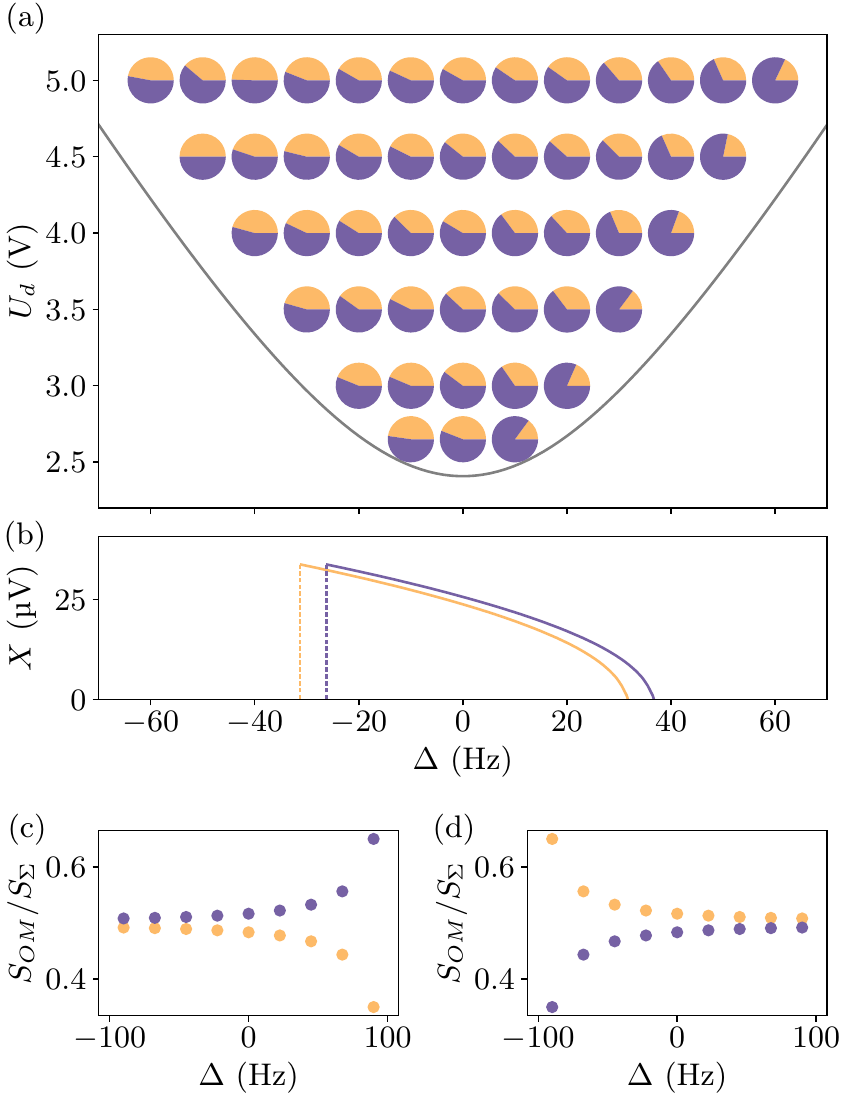}
    \caption{\textbf{Relative occupation of solutions.} (a)~Cake diagrams show the relative time the system spent in the antisymmetric (orange) and symmetric (purple) solution, at a given detuning $\Delta$ and driving $U_\mathrm{d}$. The coupling coefficient in this measurement was $J = -\SI{96(13)e6}{\hertz\squared}$. The noise strength $U_\xi$ was varied in proportion with the response amplitude of the KPOs to keep the switching rate within reasonable bounds. As a consequence, only the qualitative result at each position is significant, i.e., which solution has the longer average dwell time. A gray line indicates the boundary of the Arnold tongue. See Appendix~\ref{sec:app_E} for details. (b)~Schematic shapes of antisymmetric (orange) and symmetric (purple) solutions as a function of detuning. The splitting is exaggerated for better visibility. (c)~Normalised Onsager-Machlup action $S_\mathrm{OM}$ for switching away from antisymmetric (orange) and symmetric (purple) solutions. The results are normalised with the sum of the symmetric and antisymmetric values ($S_\Sigma$) and shown as a function of normalised detuning. A higher $S_\mathrm{OM}$ corresponds to a longer dwell time in a state (see Appendix~\ref{sec:apOM}). The values are calculated for the experimental parameters with $\lambda = 0.0004$, corresponding to $U_\mathrm{d} \approx \SI{6.48}{\volt}$. (d)~Same calculation as in (c) but with the opposite sign of the nonlinearity, $\beta > 0$.}
    \label{fig:fig4}
\end{figure}

Figure~\ref{fig:fig2}(a) offers a simple interpretation of several recent theory proposals for simulating the Ising ground state with KPO networks ~\cite{Goto_2016,Puri_2017_NC,Nigg_2017}. The proposals predict that for quasiadiabatic ramping of $U_{\mathrm{d}}$ with $\Delta < 0$, the stable solution emerging beyond the lowest instability threshold corresponds to the correct Ising ground-state solution. As we see in Fig.~\ref{fig:fig2}(a), such a protocol can effectively be reduced to finding the normal mode with the lowest eigenfrequency in the case of two coupled KPOs. We confirm the protocol in Fig.~\ref{fig:fig2}(c). Our system rings up to symmetric and antisymmetric states for $\Delta > 0$ and $\Delta < 0$, respectively. As the Ising ground state we seek is the antiferromagnetic one, the latter presents the correct solution. Note that the protocol would still work for $J > 0$, as both the Ising ground state and the normal-mode ordering would then be reversed. For strong coupling and larger networks, additional nonlinear effects can make the solution space more complex to map~\cite{Heugel_2022}. This will be the scope of future experiments.

The statistical spread of states in Fig.~\ref{fig:fig2}(b) can be understood from the competition between deterministic ordering due to the coupling term $J$ on the one hand, and stochastic (thermal) force noise on the other hand. When the drive is switched on suddenly, noise in a wide frequency range participates in this competition. Note, however, that the noise intensity is always finite, resulting in a finite statistical bias towards the solution preferred by the coupled system. Slow ramping of the drive additionally low-pass filters the noise and favors a deterministic ordering. This is why the outcome in Fig.~\ref{fig:fig2}(c) is neatly divided into two halves.

Instead of suppressing the noise, it can be interesting to enhance it in order to enable activated jumps between all states. In this way, we gain a ``stochastic sampling'' map of the solution space at particular values of $U_{\mathrm{d}}$ and $\Delta$, rather than just a single solution~\cite{heugel2022_TC}. Activated escape involves a random walk from the initial state (first quasipotential well) over a quasienergy barrier, and a deterministic decay into the opposite state (second quasipotential well)~\cite{dykman1979theory,Dykman_1993}. In analogy to a thermodynamic system, we naively expect that the optimal solution occupies the deepest quasipotential well and therefore has the longest average  dwell time $\tau$ between switches~\cite{Dykman_2018,margiani2021fluctuating}.

We test the stochastic sampling protocol in Fig.~\ref{fig:fig3}. White voltage noise with a standard deviation $U_{\xi,i}$ applied to each drive electrode enhances the force noise enough to cause activated jumps between the symmetric and antisymmetric solutions as a function of time; see Fig.~\ref{fig:fig3}(a)~\cite{margiani2021fluctuating}. Plotting such a data set as a function of the variables $u_s = (u_1 + u_2)/\sqrt{2}$ and $u_a = (u_1 - u_2)/\sqrt{2}$, where we define $x_i(t) = u_i \cos(\pi f_\mathrm{p} t) + v_i \sin(\pi f_\mathrm{p} t)$, allows us to identify symmetric and antisymmetric solutions, respectively, cf. Fig.~\ref{fig:fig3}(b).

We repeat the stochastic sampling for different values of $\Delta$ and $U_{\mathrm{d}}$, and we summarize the results in Fig.~\ref{fig:fig4}. The relative dwell times measured for the symmetric and antisymmetric states are shown in cake diagrams. Surprisingly, the driven out-of-equilibrium system favors the symmetric state in the entire range of parameters. This is in contrast to the behavior expected from the corresponding (equilibrium) Ising Hamiltonian $H_\mathrm{Ising}$, where the antisymmetric ground state would always have the highest occupation.

To investigate this enigma, we compare the switching paths that can carry the system from the symmetric to the antisymmetric state or vice versa. For small switching probabilities, the main contribution of the random walk is concentrated in a narrow channel in phase space~\cite{Chan_2008}, such that we can approximate the total switching rate from the optimal switching path $\mathbf{Y}_{\rm min}$ for each transition (see Appendix~\ref{sec:apOM}).

The theory results obtained for our device parameters agree with our experimental findings. As shown in Fig.~\ref{fig:fig4}(c), the results consistently predict longer dwell times for the symmetric configuration than for the antisymmetric one over the entire range of $\Delta$, meaning that the time required to escape the quasipotential well of the symmetric state is longer on average. The ordering of the dwell time coincides with that of the state amplitudes shown in Fig.~\ref{fig:fig4}(b). Interestingly, when repeating the calculations for a positive nonlinearity $\beta$, the result changes and the antisymmetric state is found to be more stable in general; see Fig.~\ref{fig:fig4}(d). Note that switching the sign of $\beta$ changes the ordering of the amplitudes of the normal modes in Fig.~\ref{fig:fig4}(b) as well. We conclude that for the system we study, the most stable state is always the one with the largest amplitude.

We can understand the experimental and theoretical results in a straightforward way. To switch from one normal mode to another, one of the participating resonators must switch to the opposite phase state, which can be achieved without energy transfer~\cite{Roychowdhury_2015,Frimmer_2019} but implies a momentum reversal. With all other parameters being approximately equal, a larger normal-mode oscillation amplitude corresponds to a larger momentum to be overcome by the stochastic process. For this reason, the system typically remains trapped for a longer time in the solution with the highest amplitude, as seen in Fig.~\ref{fig:fig4}.

Our experimental confirmation of Ising simulation protocols~\cite{Goto_2016,Puri_2017_NC,Nigg_2017} is a first step towards understanding large systems. For $N > 2$, the number of normal modes does not match the expected size of the Ising solution space, and it will be important to understand how the system evolves far beyond the parametric threshold as a function of detuning~\cite{Heugel_2022} and in the presence of persistent beating between solutions~\cite{strinati2020coherent}. Mapping all solutions of a complex system can be difficult with deterministic drives due to hysteresis. Stochastic sampling, as demonstrated here, may be a way to overcome this limitation, providing a direct tomography of the solution space for a given parameter set. More generally, stochastic sampling or related experiments such as simulated annealing can be useful for understanding the analogy between an out-of-equilibrium nonlinear resonator network and thermodynamic systems. As we show in this work, the connection between the two paradigms can be counterintuitive.

Finally, we include a brief outlook on performing Ising simulation protocols with quantum-coherent systems. Quantum systems are predicted to be more efficient in finding the Ising ground state than the corresponding classical system~\cite{Goto_2016}, which makes them a valuable resource for solving optimization problems~\cite{mohseni2022ising}. There, the competition between the timescales set by  decoherence on the one hand and energy exchange between the KPOs on the other hand implies that strong coupling is a crucial requirement for quantum adiabatic evolution. However, the combination of strong coupling and nonlinearity was shown to impact the Ising solution space in surprising ways, calling for careful calibration~\cite{Heugel_2022, heugel2022_TC}. For this reason, the development of quantum systems will benefit from methods such as stochastic sampling that allow visualizing the complete solution space.

\section*{Acknowledgments}

Fabrication was performed in nano@Stanford labs, which are supported by the National Science Foundation (NSF) as part of the National Nanotechnology Coordinated Infrastructure under Award No. ECCS-1542152, with support from the Defense Advanced Research Projects Agency’s Precise Robust Inertial Guidance for Munitions (PRIGM) Program, managed by Ron Polcawich and Robert Lutwak. This work was further supported by the Swiss National Science Foundation through Grants No.~CRSII5\_177198/1, No.~CRSII5\_206008/1, and No.~PP00P2~190078, and by the
Deutsche Forschungsgemeinschaft (DFG) through Project No.~449653034.. J.d.P.~acknowledges financial support from the ETH Fellowship program (Grant No.~20-2 FEL-66). O.Z. acknowledges funding from the Deutsche Forschungsgemeinschaft (DFG) Project No.~449653034.

\newpage

\appendix

\section{Devices and Basic Characterization}\label{sec:apDevices}

\begin{figure}[t]
    \includegraphics[width=\columnwidth]{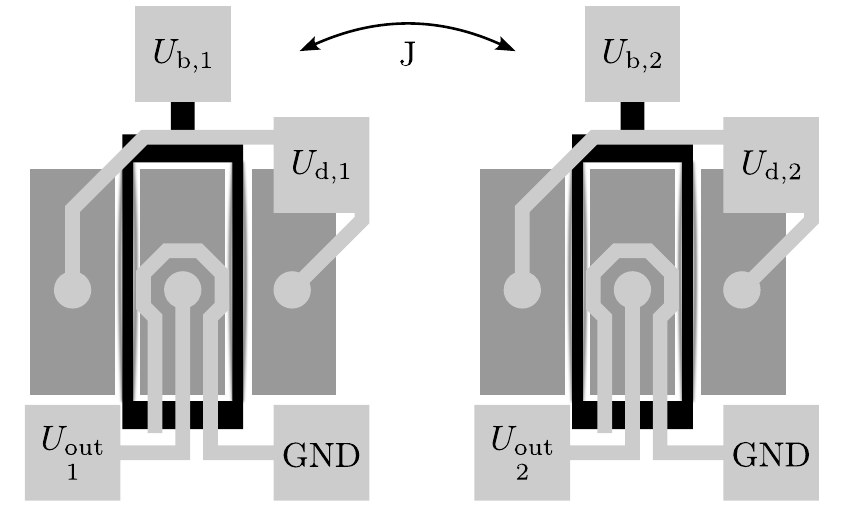} 
    \caption{\textbf{Schematic of the resonators.} The oscillating tuning forks are drawn in black. Each fork is fixed at the top and polarised by applying a bias voltage to the $U_\mathrm{b,i}$ pads (light gray). The two branches of a single fork are then driven into opposite motion by a capacitive force exerted from a varying voltage on the dark grey electrodes next to the beams. Readout is performed by measuring the voltage on the center dark gray electrode relative to ground \cite{agarwal2008study}. Connections and pads on top of the chip are shown in light gray. The two resonators are mechanically coupled via their common substrate. }
   \label{fig:SMsch}
\end{figure}

Detailed schematics showing the two tuning-fork resonators and the electrodes for capacitive driving are shown in Fig.~\ref{fig:SMsch}.

\begin{figure}[t]
    \includegraphics[width=\columnwidth]{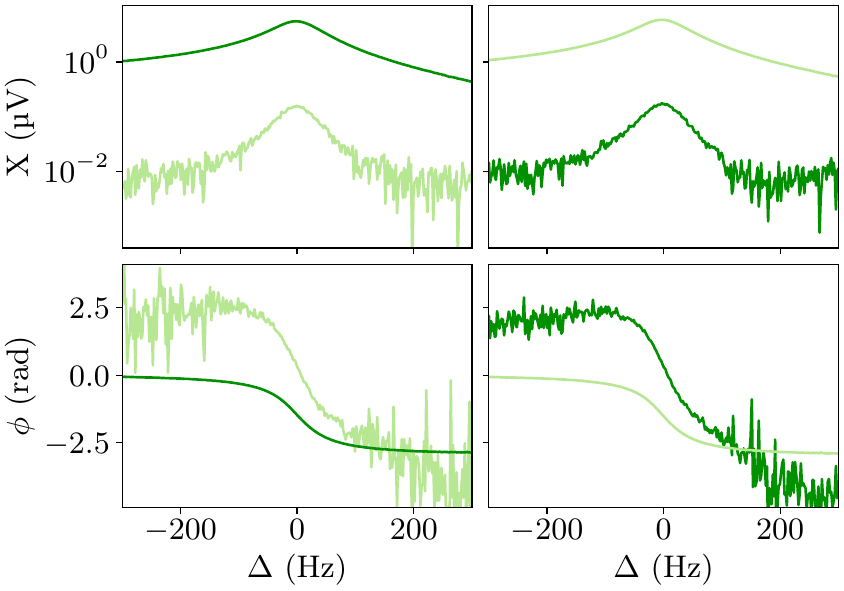} 
    \caption{\textbf{Linear response of coupled resonators.} Phase and amplitude of both resonators while only one is driven by an external force. Left: resonator 1 is driven with $U_f = \SI{20}{\milli\volt}$ and resonator 2 follows. Right: resonator 2 is driven while 1 follows. The non-driven resonator generally shows a narrower peak in its amplitude response, resulting from the combination of its own response function with the frequency-dependent force amplitude it experiences from the externally driven resonator. Similarly, the phase of the non-driven resonator changes by \SI{180}{\degree} when crossing the resonance, which is the sum of its own harmonic response phase and the phase of the driven partner.}
   \label{fig:SMcalib}
\end{figure}

The resonators' properties were extracted from fits to their driven response. Applying a small driving force to only one of the resonators allows us to extract the precise resonance frequency as well as the quality factor. The same measurement can give a clear indication of coupling between the resonators when looking at the response of the nondriven resonator, cf. Fig.~\ref{fig:SMcalib}. Due to the frequency-dependent driving amplitude and phase that the second resonator experiences, its amplitude response is narrower than usual and its phase changes by \SI{180}{\degree} when crossing the resonance.

The parametric threshold was found by measuring a full Arnold tongue, whose tip (lowest point) corresponds to the effective threshold on resonance $U_\mathrm{th}^0$ while the outer shape is determined by 
\begin{equation}\label{eq:U_th}
U_\mathrm{th} = \frac{U_\mathrm{th}^0 Q}{2}\sqrt{4\left(\left(\frac{\omega^2}{\omega_0^2} - 1\right)^2+ \frac{1}{Q^2}\right)}\,,
\end{equation}
cf. Fig.~\ref{fig:fig1}(b)~\cite{lifshitz2008nonlinear}. Equation~\eqref{eq:U_th} is a reformulation of the formula for $f_{\pm}$ shown in the caption to Fig.~\ref{fig:fig2}(a). The parametric modulation depth is calibrated as $\lambda = \frac{2 U_{d}}{Q U_\mathrm{th}^0}$. The amplitudes in a single line above threshold can then be used to determine the Duffing nonlinearity factor by fitting the parametric frequency response amplitude:

\begin{equation}
    |X(\omega)| = \Re\left(\sqrt{\frac{4 \omega_0^2}{3\beta}(\pm i\sqrt{c}+b-1)}\right)\,,
\end{equation}
where $\omega = \pi f_p$, $b = \frac{\omega^2}{\omega_0^2}$ and $c = \frac{-\lambda^2}{4} + \frac{\gamma^2 \omega^2}{\omega_0^4}$, cf. Fig.~\ref{fig:fig1}(c). Finally, we extract the coupling strength from the data in Fig.~\ref{fig:fig1}(d). With the frequency difference $\Delta_f$ between the end of the antisymmetric response and the end of the symmetric response, the coupling coefficient can be calculated as $J = 2\pi \Delta_f\omega_0$. The sign of the coupling then follows from the ordering of symmetric and antisymmetric responses. A comparable result can be extracted from the resonant amplitudes in Fig.~\ref{fig:SMcalib}. For $X_\mathrm{d}$ and $X_\mathrm{n}$ being the amplitudes at resonance of the driven and coupled, nondriven resonator respectively, we find $J = -\frac{X_\mathrm{n}}{X_\mathrm{d}}\frac{\omega_0^2}{Q}$. 

\section{Stochastic Sampling}\label{sec:app_E}
To test the stochastic sampling protocol as shown in Fig.~\ref{fig:fig4}, we measured a set of time traces for different frequencies $\Delta$ and parametric driving strengths $U_\mathrm{d}$. Each measured time trace is \SI{360}{\second} long with 3597 samples per second. We heuristically adjusted the standard deviation $\sigma_\mathrm{\xi}$ of the white noise $U_\mathrm{\xi,i}$ applied to the resonators together with the drive as $\sigma_\mathrm{\xi} \approx 4 \times 10^4 \times (1.57X-0.01)$ to reach a reasonable switching rate over the parameter range of the measurement.

In a first analysis step, each data point from a single KPO was assigned to one of the phase states labeled $0$ or $\pi$. In order to do so, we defined circles in phase space whose center point were the stable attractors and whose radii were $0.6$ times the state amplitude. A data point was assigned to a phase state when it was within the respective circle. If the data point was outside both circles, it was assigned to the same state as the previous data point. See Ref.~\cite{margiani2021fluctuating} for details.

In a second step, we then compared the relative states of both KPOs, counting how often the two phase states are equal or opposite. The results of those polls are shown in the cake diagrams in Fig.~\ref{fig:fig4}. They give a qualitative estimate of which state the system prefers to be in.

\section{Onsager-Machlup Function}\label{sec:apOM}
The switching process between stable oscillation states induced by weak-noise can be described analogously  to noise-activated jumping over a barrier $W$  in  equilibrium systems~\cite{Stambaugh_2006,Hanggi_1990}. However, the barrier $W$ between two quasistable solutions of a driven system resides in a quasipotential structure in a rotating frame. The  Onsager-Machlup formalism can be used to obtain an estimate for the barrier $W$~\cite{Lehmann_2003, Wio_2013}. The Onsager-Machlup action is defined by:
\begin{equation}\label{eq:Onsager_Machlup}
    S_{\rm OM}[\mathbf{Y}] = \int_{t_i}^{t_f} \frac{1}{4}\left(\dot{\mathbf{Y}}-\mathbf{f}(\mathbf{Y}) \right)^2 dt\,,
\end{equation}
where $t_i$ ($t_f$) is the initial (final) time of the trajectory of the $N$ resonator system, $\mathbf{Y}(t)=(u_1,v_1,...,u_N,v_N)$, which obeys the equation of motion  $\dot{\mathbf{Y}} = \mathbf{f}(\mathbf{Y})$.
For the switching rate $\Gamma$ in the weak-noise limit one can derive the scaling $\Gamma \propto \exp(-2 W/\sigma^2)$ with noise variance $\sigma^2$ and barrier $W = S_{\rm OM}[\mathbf{Y}_\mathrm{min}]$, where $\mathbf{Y}_\mathrm{min}$ is the optimal transition path minimizing $S_{\rm OM}$~\cite{Lehmann_2003, Stambaugh_2006}. We can thus conclude that the stable state with higher barrier will be more likely in an noisy environment. 

\clearpage

\bibliography{aipsamp}

\end{document}